**Continuum model of strong light-matter coupling for molecular polaritons**

Suman Gunasekaran[*1], Ryan F. Pinard[2], and Andrew J. Musser[2]

[1]Department of Chemistry, Stanford University, Stanford, California 94305

[2]Department of Chemistry and Chemical Biology, Cornell University, Ithaca, New York 14853

*E-mail: sumang@stanford.edu

## ABSTRACT

Strong coupling between light and matter generates hybrid polariton modes. We present a continuum formalism that expresses the polariton modes in terms of light and matter densities of states (DOS). We derive exact expressions for the light and matter DOS for a planar cavity containing a strongly dispersive dielectric. We show that these DOS depend exclusively on the linear response of the cavity components, i.e., the reflectance of the mirrors and the susceptibility of the dielectric. We further show that, within the strong coupling regime, the light and matter DOS are well-approximated by coupled mode theory. Altogether, our continuum formalism offers a unified treatment of polaritons that connects the framework of light-matter coupling with that of linear dispersion.

# I. INTRODUCTION

When a dielectric is placed inside an optical cavity, the interaction between the electromagnetic field of the cavity and the polarization field of the dielectric results in hybrid normal modes known as polaritons [1,2]. Polaritons are typically described in terms of light-matter coupling, where the polaritons arise from the coupling between discrete modes of light (i.e., photons) and matter (i.e., excitons, phonons, etc.) [3]. Polaritons acquire a distinct meaning in the strong light-matter coupling regime, when the polaritons modes are spectrally distinct from the originating light and matter modes [4,5]. In this regime, the hybrid nature of polaritons gives rise to unique phenomena that light and matter do not exhibit on their own [6–8].

While light-matter coupling serves as a useful framework, it should be emphasized that polaritons can be defined without invoking light-matter coupling. Polaritons represent the normal modes of the dielectric-cavity system [2,9], and light-matter coupling merely provides a perturbative solution for these normal modes [10,11]. The light-matter interaction can instead be encoded in the permittivity of the dielectric, and the polariton modes can simply be obtained from the dispersion relation [12]. Surprisingly, while both the light-matter coupling and dispersion-based approaches are used extensively in experimental modeling, there is a marked disparity in the rigor of their implementation. The dispersion-based approach, often implemented using the transfer matrix technique [4,13,14], can quantitatively model the scattering spectra of cavities, whereas the light-matter coupling approach often takes the form of a heuristic coupled oscillator model [5,15] that is only fit to the data. The comparative limitation of the light-matter coupling approach highlights a key oversight in the established understanding of polaritons and their relationship to linear dispersion.

We present a continuum formalism of polaritons that unifies the framework of light-matter coupling with that of linear dispersion. The central principle of our formalism is to describe the polaritons by a continuous spectrum, expressed in terms of light and matter local densities of states (LDOS). This approach is a notable departure from polariton orthodoxy. Though polaritons have been described in terms of DOS previously [16,17], polaritons are almost exclusively described in terms of discrete, quasi-normal modes [18,19]. We find that introducing continuous spectra dramatically simplifies the modal analysis of polaritons. We consider a classical light-matter interaction in a planar cavity, which is fully described by the linear response of the cavity components, namely by the permittivity of the dielectric and the reflectance of the mirrors. From these quantities alone, we derive exact expressions for the LDOS for light and matter, which define the polariton modes. Thus, our approach is not only theoretically rigorous, but also immensely practical, since it is based entirely on empirical quantities.

## II. CONTINUUM FORMALISM OF POLARITONS FOR A PLANAR CAVITY

For a concrete discussion of our continuum formalism, we will derive the exact light and matter LDOS for the transverse-electric (TE) modes of a planar cavity. While this is effectively a scalar, one-dimensional problem, our formalism can be readily extended to three dimensions by including additional vector notation. We consider a planar cavity oriented along the z-direction with mirrors at $z = 0$ and $z = L$ that contains a homogenous dielectric medium (Fig. 1(a)). The polariton modes of the cavity comprise both an electric field $\boldsymbol{\mathcal{E}}(x, z, t)$ and a polarization field $\boldsymbol{\mathcal{P}}(x, z, t)$. Due to the planar symmetry of the cavity, these fields can be expressed as plane waves of the form

$$\boldsymbol{\mathcal{E}}(x, z, t) = E(z) e^{i(k_\parallel x - \omega t)} \hat{\boldsymbol{y}} \tag{1}$$

and

$$\boldsymbol{\mathcal{P}}(x,z,t) = P(z)e^{i(k_{\parallel}x-\omega t)}\hat{\boldsymbol{y}} \ , \tag{2}$$

which are defined by angular frequency $\omega$ and in-plane wavevector $k_{\parallel}$ [20]. We orient $k_{\parallel}$ along the $x$-direction, without loss of generality. For TE polarization, the electric field is polarized along the $y$-direction, transverse to the direction of $k_{\parallel}$, indicated by unit vector $\hat{\boldsymbol{y}}$. The spatial profiles $E(z)$ and $P(z)$ depend on the properties of cavity as well as the parameters $\omega$ and $k_{\parallel}$.

The electric and polarization fields are governed by a set of differential equations. The electric field $\boldsymbol{\mathcal{E}}(x,z,t)$ is governed by the wave equation [21]

$$\nabla^2\boldsymbol{\mathcal{E}} - \frac{\varepsilon_\infty}{c^2}\frac{\partial^2\boldsymbol{\mathcal{E}}}{\partial t^2} = \frac{1}{\varepsilon_0 c^2}\frac{\partial^2\boldsymbol{\mathcal{P}}}{\partial t^2} \ , \tag{3}$$

where $c$ is the speed of light, $\varepsilon_0$ is the vacuum permittivity, and $\varepsilon_\infty$ is the relative permittivity of the background dielectric, which we distinguish from the strongly dispersive, active component of the dielectric that is described by polarization $\boldsymbol{\mathcal{P}}(x,z,t)$. Here we will consider the active component to comprise a total of $N$ oscillators that are uniformly distributed throughout the volume $V$ of the cavity, with number density $n = N/V$. The polarization of these oscillators is given by

$$\boldsymbol{\mathcal{P}}(x,z,t) = \sum_{j=1}^{N}\boldsymbol{\mathcal{P}}_j(x,z,t) \ , \tag{4}$$

where $\boldsymbol{\mathcal{P}}_j(x,z,t)$ is the polarization of oscillator $j$ with spatial profile $P_j(z)$. Oscillator $j$ is described by the Lorentz model [22], with resonance frequency $\omega_j$, damping rate $\gamma_j$, and oscillator strength $f_j$. The response of polarization $\boldsymbol{\mathcal{P}}_j(x,z,t)$ to applied electric field $\boldsymbol{\mathcal{E}}(x,z,t)$ is given by

$$\frac{\partial^2\boldsymbol{\mathcal{P}}_j}{\partial t^2} + \gamma_j\frac{\partial\boldsymbol{\mathcal{P}}_j}{\partial t} + \omega_j^2\boldsymbol{\mathcal{P}}_j = \frac{nq^2 f_j}{m}\boldsymbol{\mathcal{E}} \ , \tag{5}$$

where $q$ is the electron charge, and $m$ is the electron mass [21]. Note that we neglect the Clausius-Mossotti relation [23] that distinguishes the local driving field from the macroscopic field $\boldsymbol{\mathcal{E}}(x, z, t)$.

Applying the plane wave basis (Eq. (1) and Eq. (2)) to the differential equations (Eq. (3) and Eq. (5)) yields simplified equations for the spatial profiles $E(z)$ and $P_j(z)$, given as

$$E''(z) + \left(\varepsilon_\infty k_0^2 - k_\parallel^2\right)E(z) = -\frac{k_0^2}{\varepsilon_0}\sum_{j=1}^{N} P_j(z) \ , \tag{6}$$

and

$$\left(\omega_j^2 - \omega^2 - i\omega\gamma_j\right)P_j(z) = \varepsilon_0\varepsilon_\infty\Omega_j^2 E(z) \ , \tag{7}$$

where $k_0 = \omega/c$ is the vacuum wavevector, and prime notation is used for derivatives. In Eq. (7), we additionally introduce the light-matter coupling parameter

$$\Omega_j = \sqrt{\frac{nq^2 f_j}{\varepsilon_0\varepsilon_\infty m}} \ . \tag{8}$$

The right-hand sides of Eq. (6) and Eq. (7) signify the coupling between the electric field and the polarization field. Conceptually, an oscillating electric field induces an oscillating polarization field (Eq. (7)), which in turn radiates an oscillating electric field (Eq. (6)). Hence, the light-matter interaction is described by a pair of coupled equations, and the normal modes, i.e., the polariton modes, will comprise both the electric and polarization fields.

To calculate the polariton modes, the central principle of our approach is to decouple Eq. (6) and Eq. (7) by introducing self-energies, so that the light and matter components of the polariton modes can be calculated independently. This is achieved by replacing the right-hand sides of Eq. (6) and Eq. (7) with the terms $\Sigma_E(\omega, z)E(z)$ and $\Sigma_{P_j}(\omega, z)P_j(z)$, respectively, yielding the decoupled equations

$$E''(z) + \left(\varepsilon_\infty k_0^2 - k_\parallel^2\right)E(z) - \Sigma_E(\omega, z)E(z) = 0 \tag{9}$$

and

$$\left(\omega_j^2 - \omega^2 - i\omega\gamma_j\right)P_j(z) - \Sigma_{P_j}(\omega, z)P_j(z) = 0 \ . \tag{10}$$

The quantities $\Sigma_E(\omega, z)$ and $\Sigma_{P_j}(\omega, z)$ represent the self-energies that must be derived. Once the self-energies are determined, the light and matter LDOS can be obtained from the imaginary part of the Green's functions associated with Eq. (9) and Eq. (10). By calculating the LDOS from the Green's function, our approach can naturally accommodate energy loss, which is present in realistic cavities but often neglected in polariton models.

## III. LOCAL DENSITY OF STATES OF THE ELECTROMAGNETIC FIELD

To calculate the photonic LDOS for the planar cavity, the self-energy $\Sigma_E$ for the electric field must first be derived. The self-energy can be obtained by introducing the susceptibility $\chi(\omega)$, defined by the relation $P(z) = \varepsilon_0\chi(\omega)E(z)$. Following from Eq. (7), the susceptibility is given by

$$\chi(\omega) = \sum_{j=1}^{N} \chi_j(\omega) = \sum_{j=1}^{N} \frac{\varepsilon_\infty \Omega_j^2}{\omega_j^2 - \omega^2 - i\gamma_j\omega} \ , \tag{11}$$

where $\chi_j(\omega)$ is the susceptibility of $P_j(z)$, satisfying $P_j(z) = \varepsilon_0\chi_j(\omega)E(z)$. Using the susceptibility, Eq. (7) can be expressed in the form of Eq. (10) provided that the self-energy is defined to be

$$\Sigma_E(\omega, z) = -k_0^2\chi(\omega) \ . \tag{12}$$

We see that the self-energy of the electric field assumes a simple form, allowing the photonic LDOS of the cavity to be conveniently calculated.

The photonic LDOS associated with Eq. (9) can be calculated from the Green's function $g_E(z,\xi)$ that satisfies the differential equation

$$\left[\frac{\partial^2}{\partial z^2} + \left(\varepsilon_\infty k_0^2 - k_\parallel^2\right) - \Sigma_E(\omega)\right] g_E(z,\xi) = \delta(z-\xi) \ , \tag{13}$$

subject to outgoing boundary conditions [24,25]. To solve for $g_E(z,\xi)$, we note that Eq. (13) is equivalent to Eq. (9) when $z \neq \xi$. Therefore, $g_E(z,\xi)$ can be constructed from solutions of Eq. (9). The general solution to Eq. (9) can be expressed in the form $E(z) = C(Ae^{ik_\perp z} + Be^{-ik_\perp z})$ for some coefficients $A$, $B$, and $C$, where

$$k_\perp = \sqrt{\varepsilon(\omega)k_0^2 - k_\parallel^2} \tag{14}$$

is the out-of-plane wavevector, and

$$\varepsilon(\omega) = \varepsilon_\infty + \chi(\omega) \tag{15}$$

is the relative permittivity of the dielectric. We can allow $g_E(z,\xi) = E(z)$ for some fixed $\xi$ by choosing coefficients $A$ and $B$ that satisfy the outgoing boundary conditions at $z = 0$ and $z = L$ and by choosing coefficient $C$ that satisfies the continuity conditions at $z = \xi$.

To satisfy the outgoing boundary condition at $z = L$, we imagine illuminating the cavity from the left, generating a left-to-right field $E_{LR}(z)$ that only features an outgoing wave for $z > L$ (Fig. 1(a)). Within the cavity, this field takes the form $E_{LR}(z) = C_+\left(e^{ik_\perp z} + r_2 e^{i\varphi} e^{-ik_\perp z}\right)$ for some coefficient $C_+$, where $r_2$ is the reflection coefficient of the right mirror and $\varphi = 2k_\perp L$ is the round-trip phase accumulation within the cavity. Similarly, the outgoing boundary condition at $z =$ 0 is satisfied by a right-to-left field $E_{RL}(z)$, with light incident from the right, that only features an outgoing wave for $z < 0$. Within the cavity, this field can be expressed as $E_{RL}(z) = C_-\left(r_1 e^{ik_\perp z} + e^{-ik_\perp z}\right)$ for some coefficient $C_-$, where $r_1$ is the reflection coefficient of the left mirror. Therefore, we can let $g(z,\xi) = E_{LR}(z)$ for $z > \xi$ and $g(z,\xi) = E_{RL}(z)$ for $z < \xi$, provided that the coefficients $C_\pm$ satisfy the continuity conditions at $z = \xi$ (Fig. 1(a)). At $z = \xi$, Eq. (13) requires $g_E(z,\xi)$ to be continuous, with a unit jump discontinuity in its derivative with respect to $z$, namely $g_E'(\xi+\varepsilon,\xi) - g_E'(\xi-\varepsilon,\xi) = 1$ in the limit $\varepsilon \to 0$. By imposing these conditions on $C_\pm$, the Green's function is calculated to be

$$g_E(z,\xi) = \begin{cases} \dfrac{\left(e^{ik_\perp \xi} + r_2 e^{i\varphi} e^{-ik_\perp \xi}\right)}{2ik_\perp(1 - r_1 r_2 e^{i\varphi})}\left(r_1 e^{ik_\perp z} + e^{-ik_\perp z}\right) & 0 \le z \le \xi \\ \dfrac{\left(r_1 e^{ik_\perp \xi} + e^{-ik_\perp \xi}\right)}{2ik_\perp(1 - r_1 r_2 e^{i\varphi})}\left(e^{ik_\perp z} + r_2 e^{i\varphi} e^{-ik_\perp z}\right) & \xi \le z \le L \end{cases}. \qquad (16)$$

The Green's function $g_E(z,\xi)$ describes the electric field at position $z$ that is radiated by a sheet of dipoles at position $\xi$. We note that Eq. (16) can be applied to any multilayer structure by using the composite reflection coefficients for the stack of layers to the left and right of any given layer. Using $g_E(z,\xi)$, the photonic LDOS can be computed without explicitly solving for the normal modes of Eq. (9).

The connection between the Green's function and the photonic LDOS can be physically motivated. The photonic LDOS in one-dimension, which we denote $\rho_E(\omega, k_\parallel, z)$, is proportional to the power radiated by a sheet of dipoles at position $z$ with frequency $\omega$ and in-plane wavevector $k_\parallel$. The radiated power is equivalent to work done by the sheet of dipoles against its own electric field. For a sheet of dipoles at position $z$, the electric field generated at position $z$ is proportional to $g_E(z,z)$, and accordingly it can be shown that $\rho_E(\omega, k_\parallel, z)$ is given by [24]

$$\rho_E(z,\omega,k_\parallel) = -\frac{2\varepsilon_\infty k_0}{\pi c}\mathrm{Im}[g_E(z,z)] \quad . \tag{17}$$

The photonic LDOS in the cavity should be compared to its value in free space, given as $\rho_E^{1D} = \sqrt{\varepsilon_\infty}/(\pi c\cos\theta)$, where $\theta$ represents an angle relative to the z-axis that is defined by the relation $\cos\theta = \sqrt{1 - k_\parallel^2/(\varepsilon_\infty k_0^2)}$. Enhancement or suppression of $\rho_E(z,\omega,k_\parallel)$ relative to $\rho_E^{1D}$, which we quantify using the Purcell factor $F_P^{1D} = \rho_E/\rho_E^{1D}$, is responsible for the Purcell effect [26,27].

In addition to the LDOS, it is useful to compute the total density of states (DOS) $D_E(\omega,k_\parallel)$, which is obtained by integrating $\rho_E(z,\omega,k_\parallel)$ over the length of the cavity. The total DOS can be expressed as

$$D_E(\omega,k_\parallel) = -\frac{2\varepsilon_\infty k_0}{\pi c}\mathrm{Im}[G_E(\omega,k_\parallel)] \ , \tag{18}$$

where

$$G_E(\omega,k_\parallel) = \frac{L^2}{i\varphi(1 - r_1 r_2 e^{i\varphi})}\left(1 + r_1 r_2 e^{i\varphi} + i(r_1 + r_2)\frac{(1 - e^{i\varphi})}{\varphi}\right) \tag{19}$$

is the value of the integral of $g_E(z,z)$ from $z = 0$ to $z = L$. The total DOS is the generalization of the photonic Hopfield coefficient, which describes the fraction of the polariton that is light as opposed to matter. Spectroscopy of cavities, which probes the photonic fraction of the polariton, essentially measures the quantity $D_E(\omega,k_\parallel)$. We emphasize $D_E(\omega,k_\parallel)$ is exact regardless of the light-matter coupling strength, and regimes of weak, strong, and ultrastrong coupling do not need to be distinguished [28].

Our analysis demonstrates that the photonic DOS of a cavity can be efficiently described by reflectance of the mirrors, $r_1(\omega)$ and $r_2(\omega)$, and the permittivity of the dielectric, $\varepsilon(\omega)$. Since these quantities can be obtained from experimental data, our approach offers a rigorous method to model polariton experiments. As an illustrative example, we will model the polariton modes for a microcavity with silver mirrors containing a 290 nm layer of 0.1 M rhodamine 6G (R6G) dye. The s-polarized reflection coefficients of the mirrors are determined using the transfer matrix model [14], and the permittivity of R6G is calculated using $\varepsilon(\omega) = \varepsilon_\infty + \chi(\omega)$, where we assume $\varepsilon_\infty$ = 2.56 and obtain $\chi(\omega)$ from the molar absorptivity of R6G via the Kramers-Kronig relations [14,29]. With the quantities $r_1(\omega)$, $r_2(\omega)$, and $\varepsilon(\omega)$, a complete analysis of the photonic DOS of the R6G microcavity can be performed. Throughout our analysis, we present $\omega$ in units of energy $(E)$ using the conversion $E = \hbar\omega$, where $\hbar$ is the reduced Planck constant.

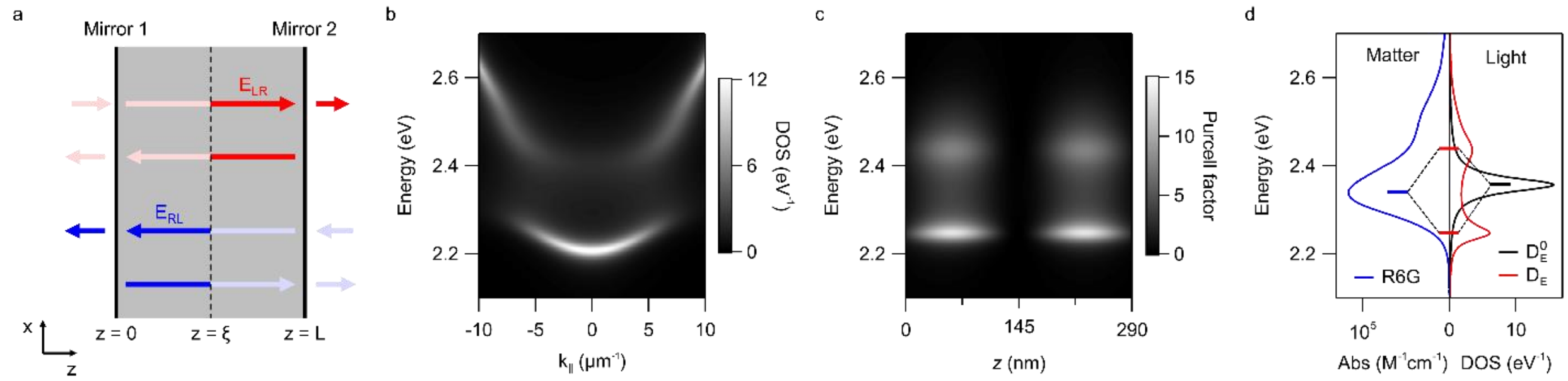

FIG. 1. (a) Schematic of the cavity structure and a depiction of $g_E(z, \xi)$, which is constructed from the scattering states $E_{LR}(z)$ (red) and $E_{RL}(z)$ (blue). The region of the dielectric is shaded in gray. (b)-(d) Analysis of the photonic DOS for the cavity structure: Ag (40 nm)/0.1 M R6G (290 nm)/Ag (200 nm). (b) The DOS $D_E(\omega, k_\parallel)$ for the TE polarization. (c) The LDOS $\rho_E(z, \omega, k_\parallel)$ corresponding to (b) at $k_\parallel$ = 5 μm$^{-1}$. (d) A comparison of $D_E(\omega, k_\parallel)$ at $k_\parallel$ = 5 μm$^{-1}$ (red), $D_E^0(\omega, k_\parallel)$ for the bare cavity (black), and the absorbance of R6G (blue). The conventional polariton hybridization diagram is overlayed.

In Fig. 1(b) we plot the DOS $D_E(\omega, k_\parallel)$ for the R6G microcavity. The photonic DOS reveals spectral bands that represent the polariton modes of the cavity. Due to the presence of R6G, the spectral bands display the characteristic avoided crossing behavior near the absorption transition of R6G (3.45 eV), resulting in the so-called upper polariton (UP) and lower polariton (LP) branches. Fig. 1(b) should be viewed as a generalization of the coupled oscillator model that is used to model polaritons. Typically, a two-level model (TLM) is implemented, and the UP and LP modes are calculated by diagonalizing a 2 × 2 matrix of the form

$$H^{(2)} = \begin{pmatrix} \omega_{ph} - i\,\gamma_{ph}/2 & V \\ V^* & \omega_{ex} - i\,\gamma_{ex}/2 \end{pmatrix} , \tag{20}$$

where $\omega_{ph}$ and $\omega_{ex}$ are the frequencies of the light and matter modes, $\gamma_{ph}$ and $\gamma_{ex}$ are the associated homogeneous linewidths, and $V$ is the light-matter coupling strength. For a given value of $k_{\parallel}$, the position and linewidth of the UP and LP peaks in $D_E(\omega, k_{\parallel})$ are a generalization of the complex eigenfrequencies of $H^{(2)}$, and the integrated area of the peaks are a generalization of the photonic component of the eigenvectors of $H^{(2)}$ (i.e., the photonic Hopfield coefficients).

The fundamental flaw of the coupled oscillator model (Eq. (20)) is that the light-matter coupling strength $V$ is treated as a fitting factor. There is no simple formula for $V$ since the light-matter interaction for a spectrally disordered collection of oscillators cannot be captured by a single number. As we demonstrate in Fig. 1(b), regardless of the spectral distribution of oscillators, the light-matter interaction is quantitatively encoded in $\chi(\omega)$, and the photonic DOS can be directly computed without any free parameters. The striking resemblance between the photonic DOS (Fig. 1(b)) and the scattering spectra that are commonly calculated using the transfer matrix model underscores the utility of the continuum approach. Essentially, one should view scattering spectra as being a reporter of the photonic DOS. Using our continuum formalism, we have elevated the coupled oscillator model to be on equal footing with the transfer matrix model.

The spatial dependence of the UP and LP modes is revealed by the LDOS $\rho_E(z, \omega, k_{\parallel})$ (Eq. (17)). In Fig. 1(c), we present $\rho_E(z, \omega, k_{\parallel})$ for $k_{\parallel} = 5$ μm$^{-1}$, which we plot in terms of Purcell factor $F_P^{1D}$. The LDOS displays an energetic splitting between UP and LP, in agreement with Fig. 1(b).

As a function of $z$, the LDOS exhibits maxima at $z = L/4$ and $z = 3L/4$, with Purcell factors exceeding 10, and a minimum at $z = L/2$, with Purcell factors under 0.1. This is the characteristic spatial profile of a $\lambda$-mode of the cavity, where $\lambda$ is the wavelength of light. In addition to the large LDOS at the UP and LP modes, there is also considerable LDOS in the avoided crossing region of the spectrum. This observation disagrees with the idealized model of collective strong-coupling, for which the avoided crossing region harbors a collection of photonically-dark matter states. As has been discussed by several authors [16,19,30–32], in realistic cavities, such as the one considered in Fig. 1(c), the supposed dark states should instead be regarded as gray states due to their partial photonic character. As a final point, we note that $F_P^{1D}$ does not directly provide the enhancement of the spontaneous emission rate of a point dipole, since $\rho_E(z, \omega, k_\parallel)$ is the one-dimensional LDOS, which describes the emission of a sheet of dipoles at $z$. The three-dimensional LDOS, describing the emission of a point dipole, can be obtained by integrating $\rho_E(z, \omega, k_\parallel)$ over all $k_\parallel$ [27].

In Fig. 1(d), we compare $D_E(\omega, k_\parallel)$ for $k_\parallel$ = 5 μm$^{-1}$ to the corresponding DOS for a bare cavity $D_E^0(\omega, k_\parallel)$, which is calculated with Eq. (18) and Eq. (19) using permittivity $\varepsilon(\omega) = \varepsilon_\infty$. The DOS of the bare cavity features a single peak (black), with an integrated area of near unity, that can be ascribed to a single photonic mode of the cavity, with a quality factor of 60. Introducing R6G into the cavity causes the DOS to split into two peaks that represent the UP and LP modes of the cavity. The total integrated area of the DOS remains near unity, but the DOS of the photonic mode is now distributed between the UP and LP modes. For this reason, the UP and LP modes are regarded as polaritonic as opposed to photonic, since the area under each peak is less than unity. The emergence of the UP and LP modes is typically rationalized using a hybridization diagram (overlayed) that depicts the perturbation to the photonic mode due to strong coupling to an optical transition of the molecule (i.e., R6G). As demonstrated by Fig. 1(d), the continuum formalism provides a quantitative description of both the density and spectral distribution of the polaritons, which cannot be adequately captured by a hybridization scheme based on discrete modes.

The hybridization scheme can be expressed rigorously within our continuum formalism using perturbation theory. The Green's function of the R6G cavity $g_E(z,\xi)$ can be expressed in terms of the Green's function of the bare cavity $g_E^0(z,\xi)$ using the Dyson equation

$$g_E(z,\xi) = g_E^0(z,\xi) + \int_0^L dz'\, g_E^0(z,z')\Sigma_{\mathrm{E}}(\mathrm{z'})g_E(z',\xi) \ , \tag{21}$$

where $g_E^0(z,\xi)$ is also defined by Eq. (16), but using permittivity $\varepsilon(\omega) = \varepsilon_\infty$. As shown by Eq. (21), the self-energy $\Sigma_E$ encodes the influence of light-matter coupling on $g_E^0(z,\xi)$ and the subsequent photonic LDOS. Thus, Eq. (21) provides a precise mathematical foundation for the polariton hybridization diagram (Fig. 1(d)).

Comparing Eq. (16) and Eq. (21) reveals two distinct, but mathematically equivalent, approaches for calculating the photonic LDOS of polaritons. In Eq. (16), the susceptibility of the dielectric is incorporated into the permittivity (Eq. (15)), and the polariton modes are a consequence of linear dispersion, whereas in Eq. (21), the susceptibility, by way of $\Sigma_E$ (Eq. (12)), is treated as a perturbation to the photonic mode, and the polariton modes arise due to strong light-matter coupling. The mathematical equivalence between Eq. (16) and Eq. (21) provides the essential connection between the frameworks of linear dispersion and light-matter coupling, and as such, our continuum formalism offers a unified view of polaritons.

## IV. LOCAL DENSITY OF STATES OF THE POLARIZATION FIELD

The matter LDOS for the planar cavity is calculated following a similar procedure to the photonic LDOS. The first step is to derive the self-energy of the polarization field $P_j(z)$. We will consider a slice of oscillator $j$ at position $z = z_m$ with polarization $P_j(z_m)$. Following from Eq. (7), the self-energy at $z = z_m$ can be formulated by expressing electric field $E(z_m)$ in terms of $P_j(z_m)$. To this end, we will invoke the relation $P_j(z) = \varepsilon_0 \chi_j(\omega) E(z)$ and remove the slice of oscillator $j$ at $z = z_m$ from $\varepsilon(\omega)$ (Eq. (15)) by subtracting the equation $L k_0^2 \chi_j(\omega) E(z) \delta(z - z_m) = (L\, k_0^2/\varepsilon_0) P_j(z_m) \delta(z - z_m)$ from Eq. (9), giving the expression

$$E''(z) + \left(\varepsilon k_0^2 - k_\parallel^2\right) E(z) - L k_0^2 \chi_j(\omega) \delta(z - z_m) E(z) = -\frac{L k_0^2}{\varepsilon_0} P_j(z_m) \delta(z - z_m). \tag{22}$$

In Eq. (22), $P_j(z_m)$ can be viewed as the source of the electric field $E(z)$. The electric field can be expressed in terms of $P_j(z_m)$ by introducing the Green's function $g_E^{(m)}(z, \xi)$ that satisfies

$$\left[\frac{\partial^2}{\partial z^2} + \left(\varepsilon k_0^2 - k_\parallel^2\right) - L k_0^2 \chi_j \delta(z - z_m)\right] g_E^{(m)}(z, \xi) = \delta(z - \xi) \ . \tag{23}$$

Comparing Eq. (23) to Eq. (22), it follows that the electric field generated by $P_j(z_m)$ is given by

$$E(z) = -\frac{L k_0^2}{\varepsilon_0} P_j(z_m) g_E^{(m)}(z, z_m) \ . \tag{24}$$

Substituting Eq. (24) into Eq. (7) reveals the self-energy at $z = z_m$ to be

$$\Sigma_{P_j}(\omega, z_m) = -\varepsilon_\infty \Omega_j^2 L k_0^2 g_E^{(m)}(z_m, z_m) \ . \tag{25}$$

An analytic expression for $\Sigma_{P_j}(\omega, z_m)$ requires a formula for $g_E^{(m)}(z_m, z_m)$. To calculate $g_E^{(m)}(z_m, z_m)$, we invoke the Dyson equation (Eq. (21)) [33], which provides the relation

$$g_E^{(m)}(z,\xi) = g_E(z,\xi) + Lk_0^2\chi_j(\omega)\int_{-\infty}^{\infty} dz'\, g_E(z,z')\delta(z'-z_m)g_E^{(m)}(z',\xi) \ , \tag{26}$$

where $g_E(z,\xi)$ represents the unperturbed Green's function that satisfies Eq. (13). Solving Eq. (26) for $g_E^{(m)}(z_m,z_m)$ yields the closed-form expression

$$g_E^{(m)}(z_m,z_m) = \frac{g_E(z_m,z_m)}{1 - Lk_0^2\chi_j(\omega)g_E(z_m,z_m)} \ , \tag{27}$$

where $g_E(z_m,z_m)$ is given by Eq. (16). Finally, after substituting Eq. (27) into Eq. (25), and dropping the dummy variable $z_m$, we derive the self-energy to be

$$\Sigma_{P_j}(\omega,z) = -\frac{\varepsilon_\infty \Omega_j^2 Lk_0^2 g_E(z,z)}{1 - Lk_0^2\chi_j(\omega)g_E(z,z)} \ . \tag{28}$$

Conceptually, $\Sigma_{P_j}(\omega,z)$ describes the influence of the electric field on $P_j(z)$ due to $P_j(z)$ itself. In Eq. (10), we see that $\mathrm{Im}\left[\Sigma_{P_j}(\omega,z)\right]$ serves the same role as $\gamma_j$ and describes the linewidth broadening, while $\mathrm{Re}\left[\Sigma_{P_j}(\omega,z)\right]$ serves the same role as $\omega_j$ and describes the level shift. When $\Omega_j$ is small, the denominator of Eq. (28) can be neglected, giving $\mathrm{Im}\left[\Sigma_P^{(j)}(\omega,z)\right] \propto \rho_E(z,\omega,k_\parallel)$ (Eq. (17)), which is the well-known the Purcell effect that directly relates $\gamma_j$ to the photonic LDOS [34].

The matter LDOS for oscillator $j$ is obtained from the Green's function $g_{P_j}(z)$ associated with Eq. (10) that satisfies the equation

$$\left[\omega_j^2 - \omega^2 - i\omega\gamma_j - \Sigma_{P_j}\right] g_{P_j}(z) = 1 \ . \tag{29}$$

It immediately follows that the Green's function is given by

$$g_{P_j}(z) = \frac{1}{\omega_j^2 - \omega^2 - i\omega\gamma_j - \Sigma_{P_j}} \quad . \tag{30}$$

Using the formula for $\Sigma_{P_j}$ (Eq. (28)), the Green's function (Eq. (30)) can be expressed in the simplified form

$$g_{P_j}(z) = g_{P_j}^0(z) + \Delta g_{P_j}(z) \ , \tag{31}$$

where $g_{P_j}^0(z) = \left(\omega_j^2 - \omega^2 - i\omega\gamma_j\right)^{-1}$ is the Green's function of the uncoupled molecule, and $\Delta g_{P_j}(z)$ is defined as

$$\Delta g_{P_j}(z) = -\frac{Lk_0^2 g_E(z,z)\chi_j(\omega)}{\omega_j^2 - \omega^2 - i\omega\gamma_j} = -\frac{k_0^2 L g_E(z,z)}{2\omega + i\gamma_j}\chi_j'(\omega) \ . \tag{32}$$

In the second part of Eq. (32), $\Delta g_{P_j}(z)$ is further simplified using the derivative of $\chi_j(\omega)$ (Eq. (11)).

The matter LDOS for molecule $j$ is related to the imaginary part of the Green's function and is given by

$$\rho_{P_j}(z,\omega,k_\parallel) = \frac{2\omega}{\pi L}\mathrm{Im}\left[g_{P_j}(z)\right] \ . \tag{33}$$

Following the notation from Eq. (31), we can express the matter LDOS in the form $\rho_{P_j} = \rho_{P_j}^0 + \Delta\rho_{P_j}$, where $\rho_{P_j}^0 = (2\omega/\pi L)\mathrm{Im}\left[g_{P_j}^0\right]$ is the LDOS of the uncoupled molecules, and the quantity

$$\Delta\rho_{P_j}(z,\omega,k_\parallel) = \frac{2\omega}{\pi L}\mathrm{Im}\left[\Delta g_{P_j}(z)\right] \ . \tag{34}$$

describes change in LDOS due to light-matter coupling. We see that not only does the matter (i.e., R6G) modify the photonic LDOS (Eq. (21)), but also the light modifies the matter LDOS (Eq.

(34)). The latter effect has garnered significant interest recently in the field of polariton chemistry [35,36].

The total LDOS of all $N$ oscillators is given by the sum $\rho_P = \sum_{j=1}^{N} \rho_{P_j}$. We are primarily interested in the change in LDOS due to light-matter coupling, which is given by

$$\Delta\rho_P(z,\omega,k_\parallel) = \sum_{j=1}^{N} \Delta\rho_{P_j} = \frac{2\omega}{\pi L}\mathrm{Im}[\Delta g_P(z)] \ . \tag{35}$$

where the quantity $\Delta g_P(z)$ is defined as

$$\Delta g_P(z) = \sum_{j=1}^{N} \Delta g_{P_j}(z) \approx -\frac{Lk_0^2}{2\omega}\chi'(\omega) g_E(z,z) \ . \tag{36}$$

In the second part of Eq. (36), the term $\gamma_j$, which is assumed to be small relative to $\omega$, is neglected in the denominator of Eq. (32), reducing the summation to that of Eq. (11). Applying the approximation of Eq. (36) to Eq. (35), we obtain a simplified expression for the change in LDOS:

$$\Delta\rho_P(z,\omega,k_\parallel) = -\frac{k_0^2}{\pi}\mathrm{Im}[\chi'(\omega) g_E(z,z)] \ . \tag{37}$$

We can additionally derive the change in DOS, denoted as $\Delta D_P(\omega,k_\parallel)$, by integrating $\Delta\rho_P(z)$ over the length of the cavity, giving the expression

$$\Delta D_P(\omega,k_\parallel) = -\frac{k_0^2}{\pi}\mathrm{Im}[\chi'(\omega) G_E(\omega,k_\parallel)] \ . \tag{38}$$

After a lengthy derivation, we have arrived at remarkably simple expressions for the change in matter LDOS (Eq. (37)) and DOS (Eq. (38)) due to light-matter coupling. These expressions provide the basis of polariton chemistry and, crucially, only depend on empirical quantities.

Therefore, our continuum formalism provides a mathematical foundation for theories of polariton chemistry that are quantitively related to experimental measurements of cavities.

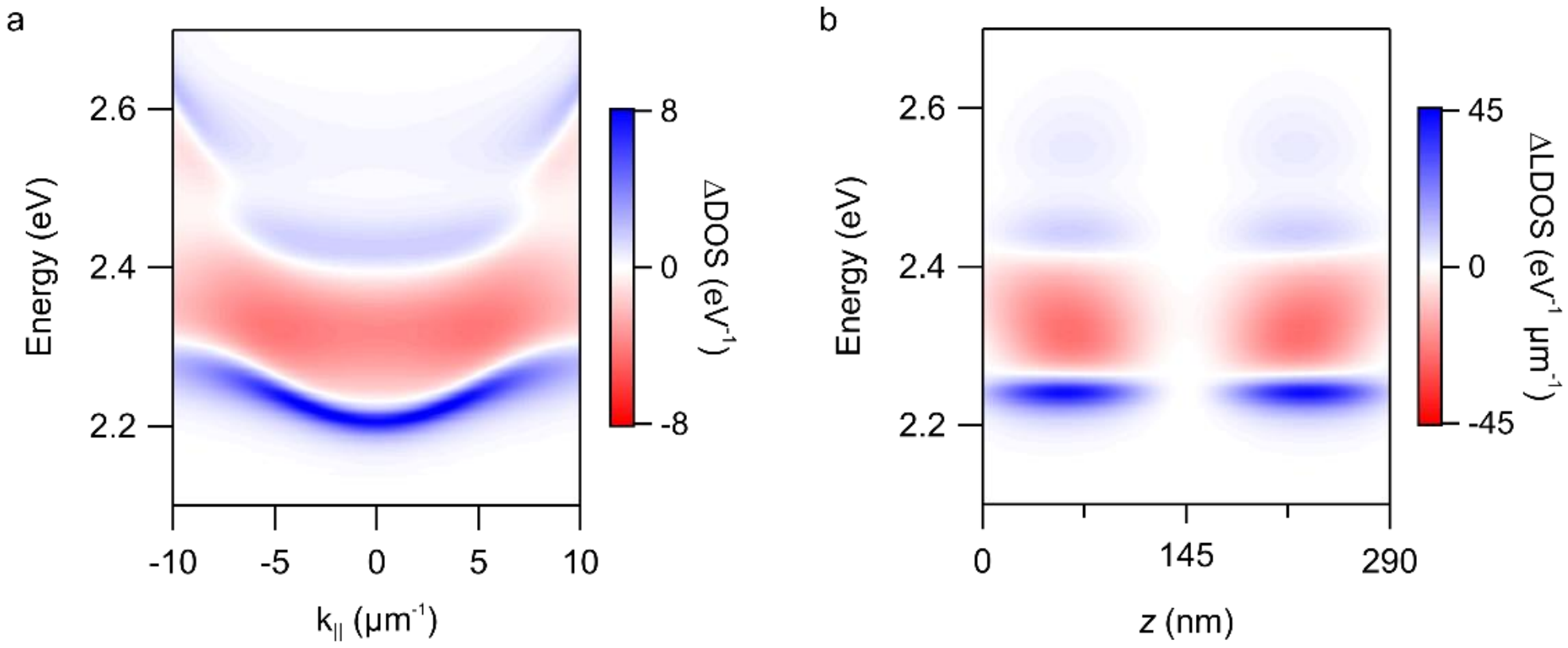

FIG. 2. Analysis of the change in matter DOS for the same cavity from Fig. 1. (a) The DOS $\Delta D_P(\omega, k_{\|})$ for the TE polarization, corresponding to Fig. 1(b). (b) The LDOS $\Delta\rho_P(z, \omega, k_{\|})$ at $k_{\|}$ = 5 μm$^{-1}$, corresponding to Fig. 1(c).

In Fig. 2, we calculate the change in matter DOS and LDOS for the R6G cavity that was considered in Figs. 1(b) and 1(c). As shown in Fig. 2(a), the change in matter DOS ($\Delta D_P(\omega, k_{\|})$) is distinct from the corresponding photonic DOS (Fig. 1(b)). While the photonic DOS of the UP mode attenuates as $k_{\|}$ approaches zero, the matter DOS reveals the opposite trend. An increase in $\Delta D_P(\omega, k_{\|})$ is observed for the UP mode as $k_{\|}$ approaches zero. This result is consistent with the light and matter Hopfield coefficients of the UP mode (i.e., the UP eigenvector of $H^{(2)}$), which behave complementarily since their sum must remain fixed. In addition to the bands of positive $\Delta D_P(\omega, k_{\|})$ for the UP and LP modes, there is a band of negative $\Delta D_P(\omega, k_{\|})$ in the avoided crossing region. The integrated area of these positive and negative spectral bands should exactly cancel out since the total DOS ($D_P(\omega, k_{\|})$) integrates to $N$ both with and without light-matter coupling. In Fig. 2(b), we see the spatial dependence of $\Delta\rho_P(z, \omega, k_{\|})$ for $k_{\|}$ = 5 μm$^{-1}$ corresponding to Fig. 1(c). As one would expect, molecules positioned at the antinodes of the

cavity mode (i.e., $z = L/4$ and $z = 3L/4$) are most significantly affected by light-matter coupling, while molecules positioned at the node of the cavity mode (i.e., $z = L/2$) are minimally affected.

Even though the molecules experience a change in DOS due to the presence of the cavity, we must emphasize that this change is negligible compared to total DOS. If we consider the cavity to have a cross-sectional area of 100 μm$^2$, the number of R6G molecules is on the order of $N \approx 10^9$. In contrast, the change in DOS is on the order of 1, since the change arises from a single photonic mode. This so-called "large $N$ problem" is the fundamental obstacle for theories of polariton chemistry [37]. One would expect any cavity-mediated effects to be exceedingly small relative to ordinary molecular processes. Moreover, not only is the integrated area of $\Delta D_P(\omega, k_\parallel)$ small relative to $N$, but also the value of $\Delta D_P(\omega, k_\parallel)$ is small relative the value of the uncoupled density ($D_P^0(\omega, k_\parallel)$) for a given frequency $\omega$. Therefore, contrary to what is suggested by the polariton hybridization diagram (Fig. 1(d)), even in the strong coupling regime, the positive peaks of $\Delta D_P(\omega, k_\parallel)$ cannot be spectrally distinguished from $D_P^0(\omega, k_\parallel)$ due to the relative density of the two quantities. This underappreciated aspect of the large $N$ problem will be discussed further in the following section.

## V. DENSITIES OF STATES WITHIN COUPLED MODE THEORY

While the light and matter LDOS can be obtained from Maxwell's equations, often one is only interested in the light and matter DOS for the whole cavity. These DOS can be calculated very conveniently and accurately using coupled mode theory (CMT) [38,39]. In CMT, light-matter coupling is treated as a perturbation to the normal modes of the uncoupled cavity-dielectric system. In the absence of light-matter coupling, the normal modes of the electric field and the polarization field, for oscillator $j$, would have time dependencies of the form $\mathcal{E}_0(t) \propto e^{-i\tilde{\omega}_0 t}$ and $\mathcal{P}_j(t) \propto$ $e^{-i\widetilde{\omega}_j t}$, with complex eigenfrequencies $\widetilde{\omega}_0 = \omega_0 - i\,\gamma_0/2$ and $\widetilde{\omega}_j = \omega_j - i\,\gamma_j/2$, respectively. Due to light-matter coupling, these time dependencies are perturbed. CMT considers the perturbation in the time-derivative of $\mathcal{E}_0(t)$ and $\mathcal{P}_j(t)$ to be proportional to the amplitude of the other field(s), given as

$$\mathcal{E}_0'(t) = -i\widetilde{\omega}_0\mathcal{E}_0(t) - i\sum_{j=1}^{N} V_j\mathcal{P}_j(t) \tag{39}$$

and

$$\mathcal{P}_j'(t) = -i\widetilde{\omega}_j\mathcal{P}_j(t) - iV_j^*\mathcal{E}_0(t)\ , \tag{40}$$

where $V_j$ represents the strength of light-matter coupling.

The coupled mode equations (Eq. (39) and Eq. (40)) can be expressed in vector notation as $i\frac{\partial}{\partial t}|\psi\rangle = H|\psi\rangle$, where $H$ is a matrix of the form

$$H = \begin{pmatrix} \widetilde{\omega}_0 & V_1 & \cdots & V_N \\ V_1^* & \widetilde{\omega}_1 & 0 & 0 \\ \vdots & 0 & \ddots & 0 \\ V_N^* & 0 & 0 & \widetilde{\omega}_N \end{pmatrix}, \tag{41}$$

and $|\psi\rangle = (\mathcal{E}_0, \mathcal{P}_1, \dots, \mathcal{P}_N)$ is a vector describing the normal mode amplitudes. Using CMT, solving Eq. (4) and Eq. (5) is reduced to an eigenvalue problem, and the polariton modes and resonance frequencies are given by the eigenvectors and eigenvalues of the matrix $H$, respectively. We note that $H$ is simply a multi-level generalization of $H^{(2)}$ (Eq. (20)).

The DOS for light and matter can be calculated within CMT from the Green's function $G(\omega)$ associated with $H$, which assumes the form of a matrix that satisfies

$$\lim_{\eta \to 0^+} [(\omega + i\eta) - H] G(\omega) = 1 \ , \tag{42}$$

where $\eta$ is a small positive number. We can introduce self-energy $\Sigma_m(\omega)$ to decouple mode $m$ from the other modes, allowing the $m^{\text{th}}$ diagonal element of Eq. (42) to be expressed as

$$[\omega - \widetilde{\omega}_m - \Sigma_m(\omega)] G_{mm}(\omega) = 1 \ , \tag{43}$$

where $G_{mm}(\omega)$ is the $m^{\text{th}}$ diagonal element of matrix $G(\omega)$. Here, we will use $m = 0$ to denote the photonic mode, and $m = j$ to denote molecule $j$. Conceptually, Eq. (43) represents an approximate form of the exact Green's functions for light (Eq. (13)) and matter (Eq. (29)). The DOS for mode $m$ is obtained from $G_{mm}(\omega)$ using the formula

$$D_m(\omega) = -\frac{1}{\pi} \mathrm{Im}[G_{mm}(\omega)] \ . \tag{44}$$

Just as in the previous sections, to calculate the DOS for light and matter, the self-energies $\Sigma_0(\omega)$ and $\Sigma_j(\omega)$ must first be derived. We note that matrix $H$ (Eq. (41)) is closely related to the Newns-Anderson impurity model [40,41], which has recently been applied to polaritons [16,42]. Following the Newns-Anderson model, the self-energy of the photonic mode, i.e., the impurity state, is given by

$$\Sigma_0(\omega) = \sum_j \frac{|V_j|^2}{\omega - \widetilde{\omega}_j} \ . \tag{45}$$

This expression for $\Sigma_0(\omega)$ closely resembles the expression for the susceptibility $\chi(\omega)$ (Eq. (11)). The susceptibility can be approximated as

$$\chi(\omega) \approx \sum_j \left( \frac{\varepsilon_\infty \Omega_j^2}{2\omega} \right) \frac{1}{\omega_j - \omega - i\,\gamma_j/2} \tag{46}$$

if we assume $|\omega_j - \omega| \ll \omega$ such that $\omega_j + \omega \approx 2\omega$, which is analogous to the rotating wave approximation [15]. Comparing Eq. (46) to Eq. (45) provides the relation

$$\Sigma_0(\omega) = -\frac{\omega F_0}{2\varepsilon_\infty}\chi(\omega) \tag{47}$$

provided that $|V_j| = \sqrt{F_0}\,\Omega_j/2$, where $F_0 \leq 1$ is the filling fraction of the photonic mode (see Appendix A). Since the filling fraction does not exceed unity, the quantity $\Omega_j$ is described as the bulk Rabi splitting, which represents an upper bound on the Rabi splitting of the cavity (i.e., $|V_j| \leq \Omega_j/2$). [43] Using $\Sigma_0(\omega)$ (Eq. (47)) in Eq. (43), the photonic Green's matrix element can be expressed as

$$G_{00}(\omega) = \left[\omega - \widetilde{\omega}_0 + \frac{\omega F_0}{2\varepsilon_\infty}\chi(\omega)\right]^{-1} , \tag{48}$$

and the photonic DOS is given by $D_0(\omega)$ (Eq. (44)). We note the self-energy $\Sigma_0(\omega)$ (Eq. (47)) differs from $\Sigma_E$ (Eq. (12)) by a factor that matches the discrepancy between the expressions for $D_E$ (Eq. (18)) and $D_0$ (Eq. (44)). Therefore, the photonic DOS calculated within CMT will be comparable to the exact value.

The matter DOS can similarly be obtained by first deriving $\Sigma_j(\omega)$. Adopting the analogous quantity from the Newns-Anderson model [40], the self-energy for oscillator $j$ is given by

$$\Sigma_j(\omega) = |V_j|^2\left(\omega - \widetilde{\omega}_0 - \Sigma_0(\omega) + \frac{|V_j|^2}{\omega - \widetilde{\omega}_j}\right)^{-1} . \tag{49}$$

Conceptually, Eq. (49) is analogous to Eq. (45), but since oscillator $j$ only couples to the photonic mode (i.e., mode 0), there is only a single term in the summation. This term invokes the effective frequency of the photonic mode, $\tilde{\omega}_0^{eff} = \tilde{\omega}_0 + \Sigma_0(\omega)$, but excludes the contribution to $\Sigma_0(\omega)$ from oscillator $j$, since oscillator $j$ cannot couple to itself. Using the expression for $\Sigma_j(\omega)$ (Eq. (49)) in Eq. (43), the matrix element $G_{jj}(\omega)$ can be expressed in the form

$$G_{jj}(\omega) = G_{jj}^0(\omega) + \Delta G_{jj}(\omega) \ , \tag{50}$$

where $G_{jj}^0(\omega) = 1/\left(\omega - \tilde{\omega}_j\right)$ is the unperturbed Green's element, and

$$\Delta G_{jj}(\omega) = \frac{\left|V_j\right|^2}{\left(\omega - \tilde{\omega}_j\right)^2} G_{00}(\omega) \tag{51}$$

is the perturbation due to light-matter coupling [40]. The DOS for oscillator $j$, denoted as $D_j(\omega)$, is given by Eq. (44). The total DOS for all molecules is the sum $D_{tot}(\omega) = \sum_{j=1}^{N} D_j(\omega)$. Accordingly, the change in total DOS due to light-matter coupling is

$$\Delta D_{tot} = -\frac{1}{\pi}\mathrm{Im}[\Delta G_{tot}(\omega)] \tag{52}$$

where

$$\Delta G_{tot}(\omega) = \sum_{j=1}^{N} \Delta G_{jj}(\omega) = -G_{00}(\omega)\Sigma_0'(\omega) \ . \tag{53}$$

In the second part of Eq. 53, we invoke the derivative of $\Sigma_0(\omega)$ (Eq. (45)) to simplify the expression [16,44]. In accordance with Eq. (47), $\Sigma_0'(\omega)$ can be approximated as

$$\Sigma_0'(\omega) \approx -\frac{\omega}{2\varepsilon_\infty}\chi'(\omega) \ , \tag{54}$$

where we consider the same approximation as in Eq. (46). Lastly, applying Eq. (54) to Eq. (52), we obtain the expression

$$\Delta D_{tot} = -\frac{\omega}{2\varepsilon_\infty \pi} \mathrm{Im}[\chi'(\omega) G_{00}(\omega)] \ . \tag{55}$$

We see that the change in matter DOS calculated within CMT (Eq. (52)) closely resembles the exact expression derived in Sec. III (Eq. (38)).

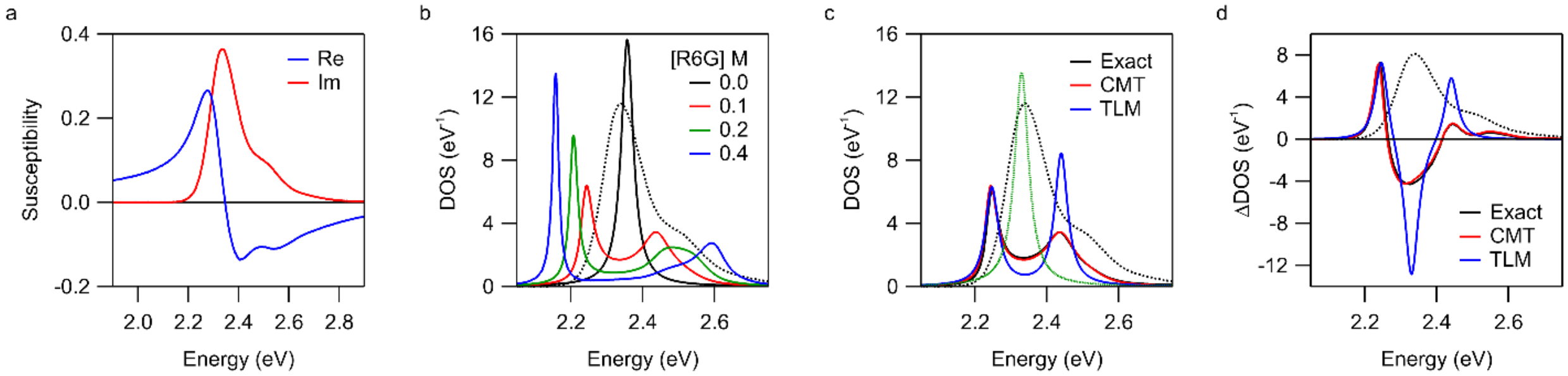

FIG. 3. (a) The real and imaginary parts of the susceptibility of R6G (0.1 M), shown in blue and red respectively. (b) The photonic DOS for a cavity with an increasing concentration of R6G. The absorbance of R6G is provided for reference (black dashed). (c) The photonic DOS for the different models. The matter state $\widetilde{\omega}_{ex}$ that is used in the TLM is shown in green. (d) The change in matter DOS for the different models, corresponding to (c).

Using CMT, the light and matter DOS can be computed without needing to consider the details of the cavity geometry. Provided that the cavity supports a spectrally isolated photonic mode, with eigenfrequency $\widetilde{\omega}_0 = \omega_0 - i\,\gamma_0/2$, the photonic DOS due to a perturbing medium with susceptibility $\chi(\omega)$ is given by Eq. (48). The filling fraction $F_0$ describes the fraction of the mode volume that is filled by the perturbing medium. While $F_0$ can be computed exactly for a specific cavity geometry (see Appendix A), $F_0$ can also be treated as a fitting parameter that serves the same role as molecular concentration.

As an example, we will use CMT to approximate the photonic DOS for the R6G cavity, which was calculated exactly in Fig. 1(d). We find that $D_E^0$ closely follows a Lorentzian line shape, and we extract $\omega_0$ and $\gamma_0$ from the center and linewidth of the peak (Fig. 1(d), black). In Fig. 3(a), we present the complex susceptibility $\chi(\omega)$ of R6G, which is derived from the molar absorptivity (Fig. 1(d)) using the Kramers-Kronig relations [14]. The photonic DOS can be calculated using Eq. (48), where we assume $F_0 \approx 1$ since the R6G fills the full cavity volume. In Fig. 3(b), the photonic DOS with and without 100 mM R6G is plotted in black and red, respectively. We find very good agreement with Fig. 1(d). We further show the dependence of the DOS on concentration, which is obtained by simply scaling $\chi(\omega)$. As expected, increasing RG6 concentration increases the splitting between the UP and LP modes. The LP mode acquires an increasingly Lorentzian lineshape as the spectral separation from the R6G absorbance (dashed black) increases. In contrast, the UP mode is inhomogeneously broadened and undergoes a complex spectral evolution with increasing concentration due to the shoulder in the R6G absorbance. This observation is consistent with the original analysis of inhomogeneous disorder, which considered idealized Gaussian broadening. [18] Using the continuum formalism, such an analysis can be performed rigorously for any absorbing dielectric.

To underscore the utility of the continuum formalism, it is instructive to compare the continuum formalism to the conventional coupled oscillator model that is used to analyze polariton modes. Fundamentally, the continuum formalism is based on the same matrix $H$ (Eq. (41)) that is used in the coupled oscillator model. Therefore, the continuum formalism does not represent a new physical model, but rather it offers a precise treatment of the same physics underlying the coupled oscillator model. To appreciate the benefits of the continuum formalism, in Fig. 3(b) we compare the DOS obtained from CMT to the TLM (i.e., $H^{(2)}$), which is the simplest implementation of the coupled oscillator model. For the TLM, the photonic DOS is calculated from $H^{(2)}$ (Eq. (20)) using Eq. (43), with $\widetilde{\omega}_{ph} = \widetilde{\omega}_0$. We emphasize that typically the output of the TLM is a discrete set of eigenfrequencies, but here we present the corresponding DOS for direct comparison with our continuum approach. As shown in Fig. 3(c), while CMT closely follows the exact DOS, the TLM deviates significantly.

The TLM suffers two key limitations. The first limitation is that the TLM neglects the inhomogeneous broadening of the matter states, which is a poor approximation for most organic molecules, such as R6G. Neglecting this broadening also introduces ambiguity into the value of $\widetilde{\omega}_{ex}$. In Fig. 3(c), the value of $\widetilde{\omega}_{ex}$ (dashed green) was chosen to best fit the DOS of the LP mode. The disagreement between $\widetilde{\omega}_{ex}$ and the R6G absorption spectrum is stark. The TLM model can be improved by introducing additional matter states into $H^{(2)}$ to reconstitute the absorption spectrum [45], but ultimately this just approximates the continuum solution (Eq. (48)).

The second, more serious limitation of the TLM is that there is no simple formula for the light-matter coupling strength $V$. Typically, $V$ is back calculated from the Rabi splitting that is measured experimentally or modelled using the transfer matrix technique. Therefore, the TLM is not predictive. Moreover, $V$ lacks a clear physical meaning since the light-matter coupling is properly expressed as a sum of terms (Eq. (45)). In Fig. 3(b), the value of $V$ was chosen to best match the Rabi splitting that was obtained quantitatively using the continuum model. While the TLM can reproduce the LP and UP peak positions, there is a large discrepancy in the DOS, particularly for the UP mode. The exact photonic DOS of the UP mode is lower and broader than is expected from the TLM due to the vibronic shoulder in the absorbance of R6G.

In the context of polariton chemistry, we are chiefly interested in the matter DOS. Crucially, spectroscopy of the cavity only probes the photonic DOS, and therefore, the matter DOS is only inferred from an assumed model. This is where the TLM can be quite misleading. The TLM suggests that the photonic DOS and matter DOS are linked, and a splitting in the photonic DOS implies a splitting in the matter DOS. We emphasize that the photonic DOS and matter DOS are distinct quantities, and one should be careful about inferring the matter DOS from the photonic DOS. Specifically, the matter Hopfield coefficient $c_{ex}$ is typically inferred by fitting the TLM to photonic DOS. We argue that $c_{ex}$ has no precise meaning, beyond simply representing the complement of the photonic Hopfield coefficient $c_{ph}$, i.e., $|c_{ex}|^2 = 1 - |c_{ph}|^2$. Since experiments only probe the photonic DOS, the assumptions underlying the validity of $c_{ex}$ are largely ignored.

Implicit in the TLM is the assumed existence of $N - 1$ dark states, which allows $H$ to be block diagonalized, with a photonic subspace given by $H^{(2)}$, yielding 2 bright states. However, as has been discussed by several authors, [16,19,45–47] due to the spatial and spectral distribution of molecules in the cavity, there are no dark states. For realistic cavities, there is no strong avoided crossing behavior, and one finds considerable photonic DOS between the UP and LP modes (Fig. 3(c)). Since the dark states and bright states are linked through the block diagonalization of $H$, if there are no purely dark states, then there are also no unique bright states. Namely, the Hopfield coefficients ($c_{ph}$ and $c_{ex}$) cannot be attributed to a single eigenstate. Instead, the photonic DOS (i.e., $c_{ph}$) is distributed over many gray states.

Since the eigenstates of $H$ cannot be partitioned into bright and dark states, we must contend with the total matter DOS $D_{tot}$. The continuum formalism reveals that $D_{tot}$ is most conveniently expressed as $D_{tot} = D_{tot}^0 + \Delta D_{tot}$, where $D_{tot}^0$ is the DOS of the uncoupled molecules, and $\Delta D_{tot}$ the change in DOS due to the cavity (Eq. (55)). In Fig. 3(d), the change in matter DOS corresponding to Fig. 3(c) is presented. While it is tempting to associate the integrated area of the positive peaks in $\Delta D_{tot}$ with the matter Hopfield coefficient $c_{ex}$, this is only possible if the positive peaks can be spectrally distinguished from $D_{tot}^0$. However, since $D_{tot}^0$ represents ~ $10^9$ molecules, the positive peaks of $\Delta D_{tot}$, with an integrated area of ~ 1, would be buried in $D_{tot}$ due to the overwhelming density of $D_{tot}^{0}$. Even far into the tails of the distribution, $D_{tot}^{0}$ is still larger than $\Delta D_{tot}$. Therefore, there would be no distinguishable LP and UP peaks in $D_{tot}$ with an integrated area that would be consistent with $c_{ex}$ from the TLM. Moreover, there are no dark states that can be subtracted from $D_{tot}$ to resolve the inconsistency. Overall, the continuum model reveals serious shortcomings of the TLM for analyzing polaritons, particularly for polariton chemistry. While these shortcomings can be remedied by introducing more matter states into the TLM, ultimately the continuum formalism provides the proper mathematical treatment of the coupled oscillator model.

## VI. CONCLUSIONS AND OUTLOOK

We introduce a continuum formalism that describes polaritons in terms of light and matter DOS. In our formalism, we decouple the electromagnetic field and the polarization field of the polariton mode using self-energies, which allows us to construct Green's functions from which the light and matter DOS can be calculated. The self-energy of the electromagnetic field is directly related to the susceptibility $\chi(\omega)$ of the dielectric. This insight provides the connection between the two seemingly distinct perspectives on polaritons of linear dispersion and light-matter coupling. From the perspective of linear dispersion, $\chi(\omega)$ is incorporated into the permittivity as $\varepsilon(\omega) = \varepsilon_\infty + \chi(\omega)$, whereas from the perspective of light-matter coupling, $\chi(\omega)$ is a perturbation to the photonic mode via the Dyson equation.

Using our continuum formalism, we derive the exact LDOS for both light and matter in a planar cavity. While the photonic LDOS of cavities has been derived in other contexts, the matter LDOS is a new quantity that we derive for polaritons. The matter LDOS is most conveniently expressed as a change in LDOS due to light-matter coupling ($\Delta\rho_P$) relative to a bare LDOS for the uncoupled molecules. Intriguingly, $\Delta\rho_P$ is related to the derivative of the susceptibility $\chi(\omega)$, meaning that $\Delta\rho_P$ can be computed from solely empirical quantities, without needing to know any microscopic details about the light-matter interaction. Therefore, the foundational principle of polariton chemistry, namely that the matter states are modified due to the cavity, can now be encapsulated in a concise and practical expression.

While the field of polariton chemistry has revealed many exciting results [48,49], it has also brought about confusion and controversy. Certain experimental observations have been difficult to reproduce [50,51], and many effects have been revealed to be measurement artifacts [52–54]. Amidst this confusion, it is critical that theorists and experimentalists have a precise, unified language for describing polariton effects. Experimentalists typically describe polaritons using the coupled oscillator model. While the coupled oscillator model provides a good approximation of the light-matter interaction, the mathematical treatment of the coupled oscillator model, based on eigenvalues of a finite matrix, is ill-suited for understanding polariton chemistry. For example, while one can often fit the TLM to experimental data, the resulting parameters, specifically the matter Hopfield coefficient $c_{ex}$, lack physical meaning. The continuum formalism provides a proper mathematical treatment of the coupled oscillator model and affords expressions for the light and matter DOS that are not only more quantitative, but also easier to implement since they depend almost exclusively on $\chi(\omega)$. Therefore, we see our continuum formalism as a useful tool for experimentalists that is grounded in a rigorous theory.

In our analysis, we have only considered the linear, classical light-matter interaction, which allows us to calculate the light and matter DOS. While this on its own does not reveal new insights into polaritonic phenomena, we believe our unique perspective on polaritons will motivate new theories for polariton chemistry. Specifically, we envision future developments of the continuum formalism to include the effect of nonlinear susceptibility [55], which will require a quantum mechanical treatment of the formalism [56]. Nonlinearities in the susceptibility are the basis of many polaritonic phenomena, such as polariton condensation [57], and may play a role in polariton chemistry.

## ACKNOWLEDGEMENTS

S.G. acknowledges funding provided by the Arnold and Mabel Beckman Foundation Postdoctoral Fellowship in Chemical Instrumentation 2023 grant (http://dx.doi.org/10.13039/100000997). R.F.P. acknowledges support from the Nexus Scholars Program at Cornell University. A.J.M. acknowledges support by the U.S. Department of Energy, Office of Science, Basic Energy Sciences, CPIMS Program under Early Career Research Program (Award No. DE-SC0021941), and the Alfred P. Sloan Foundation.

## APPENDIX A

To explicitly relate the light and matter DOS within CMT to the DOS derived from Maxwell's equations, we must consider spatial profile of the quasinormal mode of the bare cavity that was introduced heuristically as $\mathcal{E}_0(t)$ in Eq. (39) and Eq. (40). Following the notation from Eq. (1), the bare cavity supports a set quasinormal modes of the form $\boldsymbol{\mathcal{E}}_n^0(x,z,t) = E_n^0(z)e^{i(k_\parallel x-\tilde{\omega}_n t)}\hat{\boldsymbol{y}}$, where $n$ denotes the mode index. These quasinormal modes are defined by eigenfunctions $E_n^0(z)$ and eigenfrequencies $\tilde{\omega}_n = \omega_n - i\,\gamma_n/2$ that satisfy the wave equation

$$E_n''(z) + \left(\frac{\varepsilon(z)}{c^2}\tilde{\omega}_n^2 - k_\parallel^2\right)E_n(z) = 0 \ , \tag{A1}$$

subject to outgoing boundary conditions at both the left and right mirrors. In Eq. (A1), we define $\varepsilon(z) = \varepsilon_\infty$ for $0 < z < L$, and we additionally consider the permittivities and thicknesses of mirror 1 and mirror 2 (Fig. 1(a)), given by $\varepsilon(z) = \varepsilon_1(z)$ for $a < z < 0$ and $\varepsilon(z) = \varepsilon_2(z)$ for $L < z < b$, respectively. The positions $z = a$ and $z = b$ are the outer surfaces of the mirrors.

Because the cavity is an open system, the eigenfunctions afforded by Eq. (A1) do not constitute a complete orthonormal basis, precluding standard methods for calculating the matrix elements of $H$ (Eq. (41)). Following the work of Young and coworkers [58–60], we will instead compute the matrix elements of the corresponding Green's function $G(\omega)$ (Eq. (42)). To this end, we introduce the generalized norm [59,61]

$$\langle E_n | E_n \rangle = \int_a^b \varepsilon(z) E_n^2(z) dz + \frac{ic}{2\tilde{\omega}_n} \left[ \frac{\sqrt{\varepsilon_L}}{\cos\theta_L} E_n^2(a) + \frac{\sqrt{\varepsilon_L}}{\cos\theta_R} E_n^2(b) \right] , \tag{A2}$$

where $\cos\theta_{L(R)} = \sqrt{1 - k_\parallel^2 / \left(\varepsilon_{L(R)} k_0^2\right)}$, and $\varepsilon_L$ and $\varepsilon_R$ are the permittivities of the ambient media (e.g., air) to the left ($z \leq a$) and right ($z \geq b$) of the cavity, respectively. We note that $\langle E_n | E_n \rangle$ is in general complex since the integral involves $E_n^2(z)$ as opposed to $|E_n(z)|^2$, and $\langle E_n | E_n \rangle$ additionally includes surface terms at $z = a$ and $z = b$.

To calculate the photonic matrix elements of $G(\omega)$, we will revisit the Dyson equation (Eq. (21)), which expresses $g_E(z, \xi)$ of the filled cavity in terms of $g_E^0(z, \xi)$ of the bare cavity. The Green's function of the bare cavity can be expanded in terms of the quasinormal modes as [59]

$$g_E^0(z, \xi) = \sum_n \frac{c^2}{2\tilde{\omega}_n} E_n(z) E_n(\xi)\, G_{nn}^0 \ , \tag{A3}$$

where $G_{nn}^{0} = 1/(\omega - \widetilde{\omega}_n)$, and we assume $E_n(z)$ is normalized to $\langle E_n|E_n\rangle = 1$. Similarly, the Green's function of the filled cavity can be expressed as

$$g_E(z,\xi) = \sum_{n,m} \frac{c^2}{2\widetilde{\omega}_n} E_n(z)E_m(\xi)G_{nm} \tag{A4}$$

for some matrix elements $G_{nm}$. Using Eq. (A3) and Eq. (A4), the Dyson equation (Eq. (21)) can be reduced to the matrix element expression [60]

$$G_{nm} = G_{nn}^{0}\delta_{nm} + G_{nn}^{0}\sum_{q} S_{nq}G_{qm} \quad , \tag{A5}$$

where $\delta_{nm}$ denotes the Kronecker delta, and we define

$$S_{nq} = \frac{c^2}{2\widetilde{\omega}_q}\int_0^L E_n(z)E_q(z)\Sigma_{\mathrm{E}}(z)dz \quad . \tag{A6}$$

In our treatment of CMT (Eq. (41)), we consider a single photonic mode, labeled with index $n = 0$, that is described by matrix element $G_{00}$ (Eq. (48)). Using Eq. (A5), $G_{00}$ can be approximated by

$$G_{00} = G_{00}^{0} + G_{00}^{0}\sum_{q} S_{0q}G_{q0} \approx G_{00}^{0} + G_{00}^{0}S_{00}G_{00} \quad , \tag{A7}$$

assuming that photonic mode 0 is spectrally isolated from the other photonic modes $q$, allowing $S_{0q}$ for $q \neq 0$ to be neglected. We note that in the case of multimode cavities [62,63], additional terms in the summation should be included. Applying the definition of $\Sigma_{\mathrm{E}}(z)$ (Eq. (12)) to Eq. (A6), the quantity $S_{00}$ can be expressed as

$$S_{00} = -\frac{\omega^2 F_0}{2\varepsilon_\infty\widetilde{\omega}_0}\chi(\omega) \quad , \tag{A8}$$

where the filling fraction $F_n$ is defined as

$$F_n = \frac{1}{\langle E_n | E_n \rangle} \int_0^L \varepsilon_\infty E_n(z)^2 dz \quad . \tag{A9}$$

In Eq. (A9), we include $\langle E_n | E_n \rangle = 1$ to emphasize the normalization of $E_n(z)$. We can recover the original expression for $G_{00}$ (Eq. (48)) by solving for $G_{00}$ in Eq. (A7), yielding the expression

$$G_{00} = \frac{1}{\omega - \widetilde{\omega}_0 - S_{00}} \quad , \tag{A10}$$

which is equivalent to Eq. (48) if we assume that $\widetilde{\omega}_0 \approx \omega$ so that $S_{00} \approx \Sigma_0$ (Eq. (47)). Therefore, using our Green's function approach, we have explicitly connected Maxwell's equations to CMT.

From the above analysis, we see that the filling fraction bridges Maxwell's equations and CMT by endowing the matrix elements of $H$ with a physical meaning [20]. The filling fraction can be understood intuitively as the fraction of the photonic mode that is filled by the perturbing dielectric. In Eq. (A9), the photonic mode is integrated from $z = 0$ to $z = L$, which represents the bounds of the perturbing dielectric, whereas in Eq. (A2), the photonic mode is integrated from $z = a$ to $z = b$, which includes in the thickness of the mirrors. Therefore, in general we can expect the filling fraction to be less than unity due to the field penetration of the photonic mode into the mirrors. We note a slight complication that $F_n$ (Eq. (A9)) is a complex number, but $F_0$ is assumed to be real in Eq. (44). This can be viewed as an additional approximation in CMT.